\documentclass[twocolumn]{autart}
\usepackage{mathrsfs}
\usepackage{amsfonts}
\usepackage{amsmath,amssymb}
\usepackage{graphicx}
\usepackage{float}
\usepackage[dvips]{epsfig}

\def\dref#1{(\ref{#1})}

\begin{document}

\begin{frontmatter}

\title{Novel Distributed Robust Adaptive Consensus Protocols for Linear Multi-agent Systems with Directed Graphs and External Disturbances \thanksref{footnoteinfo}} % Title, preferably not more

\thanks[footnoteinfo]{This work was supported by the National Natural Science Foundation of China under grants 61104153, 11332001, 61225013,  
a Foundation for the Author of National Excellent Doctoral Dissertation of PR China,
and the State Key Laboratory of Complex System Intelligent Control and Decision. Corresponding author Zhongkui Li.}

\author[China]{Yuezu Lv}\ead{yzlv@pku.edu.cn},
\author[China]{Zhongkui Li}\ead{zhongkli@pku.edu.cn},
\author[China]{Zhisheng Duan}\ead{duanzs@pku.edu.cn},
\author[HK]{Gang Feng}\ead{megfeng@cityu.edu.hk}

         % full addresses
\address[China]{State Key Laboratory for Turbulence and Complex Systems, Department of Mechanics and Engineering Science,
College of Engineering, Peking University, Beijing 100871, China}

\address[HK]{Department of Mechanical and Biomedical Engineering, City University of Hong Kong, Hong Kong, China}


\begin{abstract}
This paper addresses the distributed consensus protocol design problem for linear multi-agent systems with directed graphs and external unmatched disturbances. A novel distributed adaptive consensus protocol is proposed to achieve leader-follower consensus for any directed graph containing a directed spanning tree with the leader as the root node.
It is noted that the adaptive protocol might suffer from a problem of undesirable parameter drift phenomenon
when bounded external disturbances exist. To deal with this issue,
a distributed robust adaptive consensus protocol is designed to guarantee the ultimate boundedness of both the consensus error and the adaptive coupling weights in the presence of external disturbances.
Both adaptive protocols are fully distributed, relying on only the agent dynamics and the relative states of neighboring agents.
\end{abstract}

\end{frontmatter}

\section{Introduction}
In recent years, the consensus problem of multi-agent systems has been an emerging research topic in the field of control,
 due to its wide applications in many areas such as satellite formation flying, cooperative unmanned systems,
and distributed reconfigurable sensor networks \cite{1}. There has been remarkable progress in achieving consensus for different scenarios; see \cite{1,2,3,4,5,6} and the references therein.
For the consensus problem, the critical task is to design distributed consensus protocols based on local information, i.e., local state or output information of each agent and its neighbors.

In this paper, we consider the consensus problem of multi-agent systems with general linear time invariant dynamics. Previous works \cite{7,8,9,10,11,12} have presented various static and dynamic consensus protocols,
which are proposed in a distributed fashion, using only
the local information of each agent and its neighbors. However, those consensus protocols involves some
design issues. To be specific, the design of those consensus protocols generally
requires the knowledge of some eigenvalue information of the Laplacian matrix associated with the
communication graph, that is, the smallest nonzero eigvenvalue
of the Laplacian matrix for undirected graphs and the smallest real part
of the nonzero eigenvalues of the Laplacian matrix for directed graphs.
Note that
the nonzero eigenvalue information of the Laplacian matrix is global information in the sense that
each agent has to know the entire communication graph to compute it.
Therefore, although these consensus protocols are proposed and can be implemented
in a distributed fashion, they cannot be designed by each agent in a distributed fashion.
In other words, those consensus protocols in \cite{7,8,9,10,11,12} are not fully distributed.

To remove the limitation of requiring global information of the communication graph,
distributed adaptive consensus protocols are reported in \cite{13,14,15,16}, which,
depending on only local information of each agent and its neighbors, are fully distributed.
It is worth noting that the adaptive protocols in \cite{13,14,15,16} are applicable to only
undirected communication graphs or leader-follower graphs where the subgraphs among followers are undirected.
Due to the asymmetry of the Laplacian matrices, it is much more difficult to design distributed adaptive consensus protocols for general directed communication graphs.
By including monotonically increasing functions to provide additional design flexibility,
a distributed adaptive consensus protocol is derived in \cite{17}
to achieve consensus for general leader-follower directed graphs containing a directed spanning tree.
The robustness of the distributed adaptive protocols with respect to uncertainties or external
disturbances is an important issue. The adaptive protocol in \cite{17}
can only be modified to be applicable to external disturbances satisfying the matching condition; see \cite{19}.
To the best of our knowledge, how to design
distributed robust adaptive consensus protocols for the case with directed graphs and general unmatched disturbances is still open.

In this paper, we aim to design distributed robust adaptive consensus protocols
for linear multi-agent systems with directed communication graphs.
A novel distributed adaptive protocol is presented and shown to achieve leader-follower consensus for directed communication graphs containing a directed spanning tree with leader as the root node. This novel adaptive protocol is fully distributed, relying on only the agent dynamics
and the relative state information of neighboring agent.
In the presence of external disturbances, it is pointed out the adaptive protocol might suffer from a problem of the parameter drift
phenomenon \cite{18}.
Therefore, the adaptive protocol is not robust in the presence of external disturbances.
To deal with this instability issue, a robust adaptive protocol
is presented, which can guarantee the ultimate boundedness of both the consensus error and the adaptive coupling weights.
The existence condition of the proposed adaptive protocols are also discussed.
Compared to the previous works \cite{17} and \cite{19}, the contribution of this paper
is at least two-fold. First, the adaptive protocol proposed in this paper
replaces the multiplicative functions in the adaptive protocol in \cite{17} by
novel additive functions. In this case, a simple quadratic-like Lyapunov function, rather
than the complicated integral-like one in \cite{17}, can be used to derive the result.
Second, in contrast to the adaptive protocol in \cite{19} which works only for the case with
disturbances satisfying the restrictive matching condition,
the robust adaptive consensus protocol given in this paper is applicable to the case of general bounded disturbances.
It should be mentioned that the methods used to derive the results in this paper
are quite different from those in \cite{17} and \cite{19}.

The rest of this paper is organized as follows. The mathematical preliminaries used in this paper is summarized in Section 2.
The distributed adaptive consensus protocol is designed in Section 3 for general directed leader-follower graphs.
A novel robust adaptive consensus protocol is presented in Section 4 to deal with external disturbances.
Simulation results are presented in Section 5. Section 6 concludes this paper.

\section{Mathematical Preliminaries}
Let $\mathbf{R}^{n\times m}$ be the set of $n\times m$ real matrices and the superscript $T$ donates transpose for real matrices. $I_N$ represents the identity matrix of dimension $N$ and $I$ denotes the identity matrix of an appropriate dimension. $\textbf{1}$ donates a column vector with all entries equal to 1. $diag(a_1,\cdots,a_N)$ represents a diagonal matrix with elements $a_i$, $i=1,\cdots,N,$ on its diagonal while $\lambda_{\min}(A)$ donates the minimal eigenvalue of $A$. The matrix inequality $A>B$ means $A$ and $B$ are symmetric matrices and $A-B$ is positive definite. $A\otimes B$ represents the Kronecker product of matrices $A$ and $B$. A nonsingular $M$-matrix $A=[a_{ij}]$ means that $a_{ij}<0$, $i\neq j,$ and all eigenvalues of $A$ have positive real parts.

A directed graph $\mathcal{G}$ consists of a node set $\mathcal{V}$ and an edge set $\mathcal{E}\subseteq \mathcal{V}\times\mathcal{V}$, in which an edge is represented by an ordered pair of distinct nodes. For an edge $(v_i,v_j)$, node $v_i$ is called the parent node, node $v_j$ the child node, and $v_i$ is a neighbor of $v_j$. A path from node $v_{i_1}$ to node $v_{i_l}$ is a sequence of ordered edges of the form ($v_{i_k}$, $v_{i_{k+1}}$), $k=1,\cdots,l-1$. A directed graph contains a directed spanning tree if there exists a node called the root such that the node has directed paths to all other nodes in the graph.

Suppose there are $N$ nodes in the directed graph $\mathcal{G}$. The adjacency matrix $\mathcal{A}=[a_{ij}]\in \mathbf{R}^{N\times N}$ of  $\mathcal{G}$ is defined by $a_{ij}=1$ if $(v_i,v_j)\in \mathcal{E}$ and 0 otherwise. The Laplacian matrix $L=[l_{ij}]\in \mathbf{R}^{N\times N}$ is defined as $l_{ii}=\sum_{j=1}^{N}a_{ij}$ and $l_{ij}=-a_{ij}$, $i\neq j$.

\newtheorem{lemma}{Lemma}
\begin{lemma}[\cite{20}]
Zero is an eigenvalue of $L$ with {\bf 1} as a right eigenvector and all nonzero eigenvalues have positive real parts.
Furthermore, zero is a simple eigenvalue of $L$ if and only if $\mathcal{G}$ has a directed spanning tree.
\end{lemma}
\begin{lemma}[\cite{21,22}]
Consider a nonsingular $M$-matrix $L$. There exists a diagonal matrix $G$ so that $G\equiv diag(g_1,\cdots,g_N)>0$ and $GL+L^TG>0$.
\end{lemma}
\begin{lemma}[\cite{23}]
If $a$ and $b$ are nonnegative real numbers and p and q are positive real numbers such that $\frac{1}{p} + \frac{1}{q} = 1$, then
$ab\leq \frac{a^p}{p} + \frac{b^q}{q}$, and the equality holds if and only if $a^p = b^q$.
\end{lemma}
\begin{lemma}[\cite{24}]
For a system $\dot{x}=f(x,t)$,
where $f(\cdot)$ is
locally Lipschitz in $x$ and piecewise continuous in $t$,
assume that there exists
a continuously differentiable function $V(x,t)$ such that
along any trajectory of the system,
%\begin{equation}\label{bou1}
$$\begin{aligned}
\alpha_1(\|x\|)\leq V(x,t)\leq \alpha_2(\|x\|),\\
\dot{V}(x,t)\leq-\alpha_3(\|x\|)+\epsilon,
\end{aligned}$$
%\end{equation}
where $\epsilon>0$ is a constant, $\alpha_1$ and $\alpha_2$ are class $\mathcal {K}_\infty$ functions,
and $\alpha_3$ is a class $\mathcal {K}$ function. %\footnote{The definitions
%of classes $\mathcal {K}$ and $\mathcal {K}_\infty$ functions
%can be found in \cite{krstic1995nonlinear}.}.
Then, the solution
$x(t)$ of $\dot{x}=f(x,t)$ is uniformly ultimately bounded.
\end{lemma}

\section{Distributed Adaptive Consensus Protocol Design}
Consider a group of $N+1$ identical agents with general linear dynamics, consisting of $N$ followers and a leader.
The dynamics of the $i$-th agent are described by
\begin{equation}\label{1c}
\dot{x}_{i}=Ax_{i}+Bu_{i},\quad  i=0,\cdots,N,
\end{equation}
where $x_i\in\mathbf{R}^n$ is the state,
$u_i\in\mathbf{R}^{p}$ is the control input,
$A$ and $B$ are constant matrices with
compatible dimensions.

Without loss of generality, let the agent in \dref{1c} indexed by 0 be the leader whose control input
is assumed to be zero, i.e., $u_0 = 0$, and the other agents be the followers. The
communication graph $\mathcal{G}$ among the $N+1$ agents is assumed to satisfy the following assumption.
\newtheorem{assumption}{Assumption}
\begin{assumption}\label{assp1}
The graph $\mathcal{G}$ contains a directed spanning tree with the leader as the root node.
\end{assumption}

Under Assumption \ref{assp1}, the Laplacian matrix $L$ associated with $\mathcal{G}$ can be partitioned as $L=\left[
\begin{array}{cc}
0& 0_{1\times N}\\
L_2&L_1
\end{array}
\right]$. In light of Lemma 1 and the definition of $M$-matrix,
it is easy to verify that $L_1\in R^{N\times N}$ is a nonsingular $M$-matrix.

%The objective of this section is to propose a novel distributed adaptive consensus protocol which can solve the leader-follower consensus problem of the agents in \dref{1c}.

The objective of this paper is to design
distributed consensus protocols such that
the $N$ agents in \dref{1c} achieve leader-follower consensus in the
sense of $\lim_{t\rightarrow \infty}\|x_i(t)- x_0(t)\|=0$,
$\forall\,i=1,\cdots,N.$

Based on the relative states of neighboring agents, we propose
a distributed adaptive consensus protocol to each follower as
\begin{equation}\label{cons1}
\begin{aligned}
&u_{i}=(c_i+\rho_i)K\xi_i,\\
&\dot{c}_i=\xi_i^T\Gamma\xi_i,\quad  i=1,\cdots,N,
\end{aligned}
\end{equation}
where $\xi_i\triangleq\sum_{j=0}^{N}a_{ij}(x_{i}-x_{j})$, $c_i(t)$ denotes the time-varying coupling weight associated with
the $i$-th follower with $c_i(0)\geq0$, $K\in\mathbf{R}^{p\times n}$ and $\Gamma\in\mathbf{R}^{n\times n}$ are the feedback gain matrices,
and $\rho_i$ are smooth functions to be determined.

Let $\xi\triangleq[\xi_{1}^{T},\cdots,\xi_{N}^{T}]^{T}$ and $x\triangleq[x_{1}^{T},\cdots,x_{N}^{T}]^{T}$. Then, we get
\begin{equation}\label{coner}
\xi=(L_1\otimes I_{N})(x-\textbf{1}\otimes x_0).
\end{equation}
Since the graph $\mathcal{G}$ satisfies Assumption \ref{assp1},
it follows from Lemma 1 that $L_1$ is a nonsingular $M$-matrix
and that the leader-follower consensus problem is solved
if and only if $\xi$ asymptotically converges to zero.
Hereafter, we refer to $\xi$ as the consensus error.
Substituting \dref{cons1} into \dref{1c} yields
\begin{equation}\label{connet}
\begin{aligned}
&\dot{\xi}=\left[I_{N}\otimes A+L_1(C+\rho)\otimes BK\right]\xi,\\
&\dot{c}_i=\xi_i^T\Gamma\xi_i,\quad  i=1,\cdots,N,
\end{aligned}
\end{equation}
where $C\triangleq diag(c_1,\cdots,c_N)$ and $\rho\triangleq diag(\rho_1,\cdots,\rho_N)$.

The following theorem provides a result on the design of the adaptive consensus protocol
(2).
\newtheorem{theorem}{Theorem}
\begin{theorem}
Suppose that the communication graph $\mathcal{G}$ satisfies Assumption \ref{assp1}.
Then, the leader-follower consensus problem of the agents in \dref{1c} can be solved under the adaptive protocol (2) with $K=-B^TP^{-1}$, $\Gamma=P^{-1}BB^TP^{-1}$,
and $\rho_i=\xi_i^TP^{-1}\xi_i$, where $P>0$ is a solution to the following linear matrix inequality (LMI):
\begin{equation}\label{alg1}
PA^T+AP-2BB^T<0.
\end{equation}
Moreover, each coupling weight $c_{i}$ converges to some finite steady-state value.
\end{theorem}
\def\proof{\noindent{\bf Proof} }
\begin{proof}
Consider the Lyapunov function candidate:
\begin{equation}\label{lya1}
V_{1}=\sum_{i=1}^{N}\frac{1}{2}g_i(2c_i+\rho_i)\rho_i+\frac{\lambda_0}{2}\sum_{i=1}^{N}\tilde{c}_i^2,
\end{equation}
where $G\triangleq diag(g_1,\cdots,g_N)$ is a positive definite matrix such that $GL_1+L_1^TG>0$, $\lambda_0$ denotes
the smallest eigenvalue of $GL_1+L_1^TG$, and $\tilde{c}_i\triangleq c_{i}-\alpha$, where $\alpha$ is a positive constant to be determined later.
It follows from Assumption 1 and Lemma 1 that $L_1$ is a nonsingular $M$-matrix. Thus
we know from Lemma 2 that such a positive definite matrix $G$ does exist.
Since $c_i(0)>0$, it follows from $\dot{c}_i\geq 0$ that $c_i(t)>0$ for any $t>0$.
Then, it is not difficult to see that $V_1$ is positive definite.

The time derivative of $V_{1}$ along
the trajectory of \dref{connet} is given by
\begin{equation}\label{lya2}
\begin{aligned}
\dot{V}_1&=\sum_{i=1}^{N}[2g_i(c_i+\rho_i)\xi_i^TP^{-1}\dot{\xi}_i+g_i\rho_i\dot{c}_i]
\\&\quad+\lambda_0\sum_{i=1}^{N}(c_i-\alpha)\dot{c}_i\\
&=\xi^T[(C+\rho)G\otimes (P^{-1}A+A^TP^{-1})\\
&\quad-(C+\rho)(GL_1+L_1^TG)(C+\rho)\otimes \Gamma]\xi\\
&\quad+\xi^T(\rho G\otimes \Gamma)\xi+\xi^T[\lambda_0(C-\alpha I)\otimes \Gamma]\xi\\
&\leq \xi^T[(C+\rho)G\otimes (P^{-1}A+A^TP^{-1})\\&\quad-\lambda_0(C+\rho)^2\otimes \Gamma]\xi+\xi^T(\rho G\otimes \Gamma)\xi\\
&\quad+\xi^T[\lambda_0(C-\alpha I)\otimes \Gamma]\xi.
\end{aligned}
\end{equation}
By using Lemma 3, we can get that
\begin{equation}\label{lya3}
\begin{aligned}
\xi^T(\rho G\otimes \Gamma)\xi\leq\xi^T(\frac{\lambda_0}{2}\rho^2\otimes \Gamma)\xi+\xi^T(\frac{1}{2\lambda_0}G^2\otimes \Gamma)\xi,
\end{aligned}
\end{equation}
and
\begin{equation}\label{lya4}
\begin{aligned}
\xi^T(\lambda_0C\otimes \Gamma)\xi\leq\xi^T(\frac{\lambda_0}{2}C^2\otimes \Gamma)\xi+\xi^T(\frac{\lambda_0}{2}I\otimes \Gamma)\xi.
\end{aligned}
\end{equation}
%Noting that $c_i>0$ and $\rho_i>0$.
Substituting \dref{lya3} and \dref{lya4} into \dref{lya2} yields
\begin{equation}\label{lya5}
\begin{aligned}
\dot{V}_1\leq&\xi^T[(C+\rho)G\otimes (P^{-1}A+A^TP^{-1})\\
&-(\frac{\lambda_0}{2}(C+\rho)^2+\lambda_0\alpha I-\frac{\lambda_0}{2}-\frac{1}{2\lambda_0}G^2)\otimes \Gamma]\xi.
\end{aligned}
\end{equation}
Choose $\alpha\geq \hat{\alpha}+\frac{\lambda_0}{2}+\max_{i=1,\cdots,N}{\frac{g_i^2}{2\lambda_0^2}}$, where $\hat{\alpha}>0$ will be determined later. Then, it follows from \dref{lya5} that
\begin{equation}\label{lya6}
\begin{aligned}
\dot{V_{1}}&\leq \xi^T[(C+\rho)G\otimes (P^{-1}A+A^TP^{-1})\\
&\quad-(\frac{\lambda_0}{2}(C+\rho)^2+\lambda_0\hat{\alpha} I)\otimes \Gamma]\xi\\
&\leq \xi^T[(C+\rho)G\otimes (P^{-1}A+A^TP^{-1})\\
&\quad-\sqrt{2\hat{\alpha}}\lambda_0(C+\rho)\otimes \Gamma]\xi.\\
\end{aligned}
\end{equation}
Let $\tilde{\xi}=(\sqrt{(C+\rho)G}\otimes P^{-1})\xi$ and choose $\hat{\alpha}$ to be sufficiently large such that $\sqrt{2\hat{\alpha}}\lambda_0G^{-1}\geq2I$. Then we can get from \dref{lya6} that
\begin{equation}\label{lya7}
\begin{aligned}
\dot{V_{1}}&\leq \tilde{\xi}^T(I_N\otimes (AP+PA^T-2BB^T))\tilde{\xi}\\
&\leq 0,
\end{aligned}
\end{equation}
where the last inequality comes directly from the LMI \dref{alg1}.
Therefore, we can get that $V_1(t)$ is bounded and so is each $c_i$. Noting that $\dot{c}_i\geq 0$, we can know that each coupling weight $c_i$ converges to some finite value. Noting that $\dot{V}_1\equiv0$ is equivalent to $\tilde{\xi}\equiv0$ and thereby $\xi\equiv0$. By LaSalle's Invariance principle \cite{25}, it follows that the consensus error $\xi$ asymptotically converges to zero. That is, the consensus problem is solved.
\end{proof}
\newtheorem{remark}{Remark}
\begin{remark}
Contrary to the consensus protocols in \cite{7,8,10,11} which use the nonzero eigenvalues of the Laplacian matrix,
the design of the proposed adaptive protocol \dref{cons1} relies on only the agent dynamics and the relative states of neighbors, which
can be conducted by each agent in a fully distributed way.
As shown in \cite{7}, a necessary and sufficient condition for the existence of the solution $P>0$ to the LMI \dref{alg1} is
that $(A, B)$ is stabilizable. Therefore, the existence condition of an adaptive protocol \dref{cons1} satisfying Theorem 1
is that $(A, B)$ is stablizable.
\end{remark}

%\begin{remark}
%%This is in contrast to , the adaptive protocol (2) is only related to . Compared to the adaptive protocols in \cite{13,14,15,16} which require the communication graph is undirected, the protocol (2) applies to directed leader-follower communication graphs.
%\end{remark}

\begin{remark}
In contrast to the distributed adaptive protocols in \cite{13,14,15,16} which are applicable to only undirected graphs,
the proposed adaptive protocol \dref{cons1} works for the case with
general directed graphs satisfying Assumption 1.
It is worth mentioning that similar distributed adaptive protocols were designed in the previous works \cite{17} and \cite{19}
for directed graphs satisfying Assumption 1.
In comparison to the adaptive protocols in \cite{17} and \cite{19}, the novel adaptive protocol \dref{cons1}
has two distinct features. First,
different from the adaptive protocol in \cite{17} which uses multiplicative functions to provide additional design flexibility,
the adaptive protocol \dref{cons1} introduces additive functions $\rho_i$ instead. As a consequence,
a simple quadratic-like Lyapunov function as in
\dref{lya1}, instead of the complicated integral-like Lyapunov function in \cite{17},
can be used to show Theorem 1.
Second, contrary to the adaptive protocol in \cite{19} which can only deal with external disturbances satisfying the
restrictive matching condition, the proposed adaptive protocol \dref{cons1}
can be modified to be applicable to general bounded external disturbances, which will be detailed in the following section.
\end{remark}

\section{Distributed Robust Adaptive Consensus Protocols}
Theorem 1 in the previous section shows that
the adaptive protocol \dref{cons1} is applicable to
any directed graph satisfying Assumption \ref{assp1} for the
case without external disturbances. In many circumstances,
the agents might be subject to various external disturbances,
for which case it is necessary and interesting to investigate whether the adaptive protocol
\dref{cons1} is robust.

The dynamics of the $i$-th agent are described by
\begin{equation}\label{1cd}
\dot{x}_{i}=Ax_{i}+Bu_{i}+\omega_i,\quad  i=0,\cdots,N,
\end{equation}
where $\omega_i\in\mathbf{R}^n$
denotes external disturbances associated with the $i$-th agent, which satisfy the following assumption.
\begin{assumption}\label{assp2}
There exist positive constants $\upsilon_i$ such that
$\|\omega_i\|\leq\upsilon_i$, $i=1,\cdots,N$, and $\|Bu_0+\omega_0\|\leq v_0$.
\end{assumption}

Note that due to the existence of
disturbances $\omega_i$ in \dref{1cd}, the relative states will not converge to zero any more
but rather can only be expected to converge
into some small neighborhood of the origin.
Since the derivatives of the adaptive gains $c_i$ in \dref{cons1}
are of nonnegative quadratic forms
in terms of the relative states, in this case it is easy to see from \dref{cons1} that
$c_{i}$ will keep growing to infinity,
which is called the parameter drift phenomenon in the classic adaptive
control literature \cite{18}.
Therefore, the adaptive protocol
\dref{cons1} is not robust
in the presence of external disturbances.

In the following, we aim to make modification on \dref{cons1} to propose a distributed robust adaptive protocol
which can guarantee the ultimate boundedness
of the consensus error and adaptive weights for the agents in \dref{1cd}.
We propose a new robust distributed adaptive consensus protocol as follows:
\begin{equation}\label{cons2}
\begin{aligned}
&u_{i}=(d_i+\rho_i)K\xi_i,\\
&\dot{d}_i=-\varphi_i(d_i-1)^2+\xi_i^T\Gamma\xi_i,\quad  i=1,\cdots,N,
\end{aligned}
\end{equation}
where $d_i(t)$ denotes the
time-varying coupling weight associated with the $i$-th follower
with $d_i(0)\geq1$,
$\varphi_i$, $i=1,\cdots,N$, are small positive constants, and the rest of the variables are defined as in \dref{cons1}.

Substituting \dref{cons2} into \dref{1cd}, it follows that
\begin{equation}\label{connet2}
\begin{aligned}
&\dot{\xi}=\left[I_{N}\otimes A+L_1(D+\rho)\otimes BK\right]\xi+(L_1\otimes I_n)\omega\\
&\dot{d}_i=-\varphi_i(d_i-1)^2+\xi_i^T\Gamma\xi_i,\quad  i=1,\cdots,N,
\end{aligned}
\end{equation}
where $C\triangleq diag(c_1,\cdots,c_N)$,
$\omega \triangleq[\omega_1^T-(Bu_0+\omega_0)^T,\cdots,\omega_N^T-(Bu_0+\omega_0)^T]^T$,
and the rest of the variables are defined as in \dref{connet}.

In light of Assumption 2, we have that
\begin{equation}\label{ome}
\|\omega\|\leq\sqrt{\sum_{i=1}^{N}(v_i+v_0)^2}.
\end{equation}
Note that $d_i(0)\geq1$ and $\dot{d}_i\geq 0$ when $d_i=1$ in \dref{connet2}.
Then, it is not difficult to see that $d_i(t)\geq 1$ for any $t>0$.

The following theorem presents a result on design of the robust adaptive consensus protocol \dref{cons2}.

\begin{theorem}
Suppose that Assumptions 1 and 2 hold. Then, both the consensus error $\xi$ and the coupling weights $d_i$, $i=1,\cdots,N$, in \dref{connet2} are uniformly ultimately bounded under the adaptive protocol \dref{cons2}
with $K=-B^TQ^{-1}$, $\Gamma=Q^{-1}BB^TQ^{-1}$, and $\rho_i=\xi_i^TQ^{-1}\xi_i$, where $Q>0$ is a solution to the LMI:
\begin{equation}\label{alg2}
AQ+QA^{T}+\varepsilon Q-2BB^T<0,
\end{equation}
where $\varepsilon>1$. The upper bound of the consensus error $\xi$ will be given in the proof.
\end{theorem}

\def\proof{\noindent{\bf Proof} }
\begin{proof}
Consider the Lyapunov function candidate:
\begin{equation}\label{lyas1}
V_{2}=\sum_{i=1}^{N}\frac{1}{2}g_i(2d_i+\rho_i)\rho_i+\frac{\lambda_0}{2}\sum_{i=1}^{N}\tilde{d}_i^2,
\end{equation}
where $\tilde{d}_i\triangleq d_{i}-\alpha$, where $\alpha$ is a positive constant to be determined later,
and the rest of the variables are defined as in \dref{lya1}. Since $g_i>0$,
$d_i(t)\geq 1$ for any $t>0$, and $\rho_i\geq0$, it can be similarly shown as in the proof of Theorem 1
that $V_2$ is positive definite.

By following similar steps in deriving Theorem 1, we can obtain the time derivative of $V_{2}$
along \dref{connet2} as
\begin{equation}\label{lyas2}
\begin{aligned}
\dot{V}_2
&\leq \xi^T[(D+\rho)G\otimes (Q^{-1}A+A^TQ^{-1}-\Gamma)]\xi\\&\quad+2\xi^T[(D+\rho)GL_1\otimes Q^{-1}]\omega\\
&\quad-\xi^T[\varphi(D-I)^2G\otimes Q^{-1}]\xi\\
&\quad-\sum_{i=1}^{N}\lambda_0\varphi_i(d_i-\alpha)(d_i-1)^2,
\end{aligned}
\end{equation}
where $\alpha$ is chosen to be sufficiently large
as in the proof of Theorem 1 and
$\varphi\triangleq diag(\varphi_1,...,\varphi_N)$.

By choosing $\alpha>1$ and using Lemma 3, we can get that
\begin{equation}\label{lyas3}
\begin{aligned}
&-(d_i-\alpha)(d_i-1)^2\\
&\quad=-(d_i-1)^3+(\alpha-1)(d_i-1)^2\\
&\quad=-(d_i-1)^3+[(\frac{3}{4})^{\frac{2}{3}}(d_i-1)^2][(\frac{4}{3})^{\frac{2}{3}}(\alpha-1)]\\
&\quad\leq-(d_i-1)^3+\frac{1}{2}(d_i-1)^3+\frac{16}{27}(\alpha-1)^3\\
&\quad=-\frac{1}{2}(d_i-1)^3+\frac{16}{27}(\alpha-1)^3.
\end{aligned}
\end{equation}
Note that
\begin{equation} \label{lyas4}
\begin{aligned}
&2\xi^T[(D+\rho)GL_1\otimes Q^{-1}]\omega\\
&=2\xi^T[(D-I)\sqrt{\varphi G}\otimes \sqrt{Q^{-1}}](\sqrt{\varphi^{-1}G}L_1\otimes \sqrt{Q^{-1}})\omega\\
&\quad+2\xi^T(\frac{1}{\sqrt{2}}\sqrt{G}\otimes \sqrt{Q^{-1}})(\sqrt{2G}L_1\otimes \sqrt{Q^{-1}})\omega\\
&\quad+2\xi^T(\frac{1}{\sqrt{2}}\sqrt{\rho G}\otimes \sqrt{Q^{-1}})(\sqrt{2\rho G}L_1\otimes \sqrt{Q^{-1}})\omega\\
&\leq \xi^T[(D-I)^2\varphi G\otimes Q^{-1}]\xi+\|(\sqrt{\varphi^{-1}G}L_1\otimes \sqrt{Q^{-1}})\omega\|^2\\
&\quad+\frac{1}{2}\xi^T(G\otimes Q^{-1})\xi+2\|(\sqrt{G}L_1\otimes \sqrt{Q^{-1}})\omega\|^2\\
&\quad+\frac{1}{2}\xi^T(\rho G\otimes Q^{-1})\xi+2\|(\sqrt{\rho G}L_1\otimes \sqrt{Q^{-1}})\omega\|^2,
\end{aligned}
\end{equation}
where we have used Lemma 3 several times to get the last inequality, and
\begin{equation} \label{lyas5}
\begin{aligned}
&2\|(\sqrt{\rho G}L_1\otimes \sqrt{Q^{-1}})\omega\|^2\\
&\quad\leq 2\|\sqrt{\rho \sqrt{G}}\otimes I_n\|^2\|(G^{\frac{1}{4}}L_1\otimes \sqrt{Q^{-1}})\omega\|^2\\
&\quad\leq \frac{1}{2}\|\sqrt{\rho \sqrt{G}}\otimes I_n\|^4+2\|(G^{\frac{1}{4}}L_1\otimes \sqrt{Q^{-1}})\omega\|^4\\
&\quad\leq \frac{1}{2}\xi^T(\rho G\otimes Q^{-1})\xi+2\|(G^{\frac{1}{4}}L_1\otimes \sqrt{Q^{-1}})\omega\|^4,
\end{aligned}
\end{equation}
where we have used matrix norm properties to get the first inequality, and Lemma 3 to get the second inequality, and to get the last inequality
we have used the fact that
$$\begin{aligned}
\|\sqrt{\rho\sqrt{G}}\otimes I_n\|^4&=\max_{i=1,\cdots,N}(\sqrt{\rho_i\sqrt{g_i}})^4\\
&\leq\sum_{i=1}^N(\sqrt{\rho_i\sqrt{g_i}})^4=\xi^T(\rho G\otimes Q^{-1})\xi.
\end{aligned}$$
Substituting \dref{lyas3}, \dref{lyas4}, and \dref{lyas5} into \dref{lyas2} yields
\begin{equation}\label{lyas6}
\begin{aligned}
\dot{V}_2&\leq \xi^T[(D+\rho)G\otimes (Q^{-1}A+A^TQ^{-1}-\Gamma)]\xi\\&\quad+\xi^T[(\rho+\frac{1}{2})G\otimes Q^{-1}]\xi-\sum_{i=1}^{N}\frac{\lambda_0}{2}\varphi_i(d_i-1)^3+\Pi_1\\
&\leq W(\xi)-\frac{1}{2}\xi^T(G\otimes Q^{-1})\xi-\sum_{i=1}^{N}\frac{\lambda_0}{2}\varphi_i(d_i-1)^3+\Pi_1,
\end{aligned}
\end{equation}
where we have used the fact that $D\geq I$,
$$\begin{aligned}
\Pi_1&=\sum_{i=1}^{N}\frac{16\lambda_0}{27}\varphi_i(\alpha-1)^3+\|(\sqrt{\varphi^{-1}G}L_1\otimes \sqrt{Q^{-1}})\omega\|^2\\&\quad +2\|(\sqrt{G}L_1\otimes \sqrt{Q^{-1}})\omega\|^2+2\|(G^{\frac{1}{4}}L_1\otimes \sqrt{Q^{-1}})\omega\|^4.
\end{aligned}$$
and
\begin{equation*}
\begin{aligned}
W(\xi)&\triangleq\xi^T[(D+\rho)G\otimes(Q^{-1}A+A^TQ^{-1}+Q^{-1}-\Gamma)]\xi\\
&\leq 0.
\end{aligned}
\end{equation*}

Note that for any positive $\delta$, we have the following assertion:
\begin{equation}\label{lyas7}
\begin{aligned}
&\frac{\delta\lambda_0}{2}(d_i-\alpha)^2\leq\frac{\delta\lambda_0}{2}(d_i-1)^2+\frac{\delta\lambda_0}{2}(\alpha-1)^2\\
&\quad=[(\frac{3\lambda_0\varphi_i}{4})^{\frac{2}{3}}(d_i-1)^2][(\frac{2\lambda_0}{9\varphi_i^2})^{\frac{1}{3}}\delta]
+\frac{\delta\lambda_0}{2}(d_i-\alpha)^2\\
&\quad\leq\frac{\lambda_0}{2}\varphi_i(d_i-1)^3+\frac{2\lambda_0\delta^3}{27\varphi_i^2}+\frac{\delta\lambda_0}{2}(\alpha-1)^2,
\end{aligned}
\end{equation}
where we have used the fact that $d_i>1$ and $\alpha>1$ to get the first inequality, and Lemma 3 to get the last inequality.
From \dref{lyas6} and \dref{lyas7}, we can obtain that
\begin{equation}\label{lyas8}
\begin{aligned}
\dot{V}_2&\leq-\delta V_2+\delta V_2+W(\xi)-\frac{1}{2}\xi^T(G\otimes Q^{-1})\xi\\
&\quad-\sum_{i=1}^{N}\frac{\lambda_0}{2}\varphi_i(d_i-1)^3+\Pi_1\\
&\leq-\delta V_2+\bar{W}(\xi)-\frac{1}{2}\xi^T(G\otimes Q^{-1})\xi+\Pi,
\end{aligned}
\end{equation}
where
$$\Pi=\sum_{i=1}^{N}[\frac{2\lambda_0\delta^3}{27\varphi_i^2}+\frac{\delta\lambda_0}{2}(\alpha-1)^2]+\Pi_1,$$
and
\begin{equation*}
\begin{aligned}
\bar{W}&=\xi^T[(D+\rho)G\otimes(Q^{-1}A+A^TQ^{-1}\\
&\quad+(1+\delta)Q^{-1}-Q^{-1}BB^TQ^{-1})]\xi,
\end{aligned}
\end{equation*}
By choosing $\delta$ such that $\varepsilon\geq1+\delta$, we can obtain that $\bar{W}(\xi)\leq0$.
Then, it follows from \dref{lyas8} that
\begin{equation}\label{lyas9}
\dot{V}_2\leq-\delta V_2-\frac{1}{2}\xi^T(G\otimes Q^{-1})\xi+\Pi.
\end{equation}

In light of Lemma 4, we can conclude from \dref{lyas9} that
both the consensus error $\xi$ and the adaptive gains $d_i$ are uniformly ultimately bounded.
Further, from \dref{lyas9}, we can get that $\dot{V}_2\leq-\delta V_2$ if
$\|\xi\|^2\geq\frac{2\Pi}{\lambda_{\min}(Q^{-1})\underset{i=1,\cdots,N}{\min} g_i}$.
Therefore, $\xi$ converges to the set
\begin{equation}\label{upperbound}
D_1=\left\{\xi:\|\xi\|^2\leq\frac{2\Pi}{\lambda_{\min}(Q^{-1})\underset{i=1,\cdots,N}{\min} g_i}\right\}
\end{equation}
with a convergence rate faster than $e^{-\delta t}$.
\end{proof}

\begin{remark}
As shown in Proposition 1 in \cite{8},
there exists a $Q>0$ satisfying \dref{alg2} if and only
if $(A,B)$ is controllable. Thus, a sufficient
condition for the existence of \dref{cons2} satisfying Theorem 2
is that $(A,B)$ is controllable, which, compared to the existence condition
of \dref{cons1} satisfying Theorem 1,
is more stringent.
Theorem 2 shows that the modified adaptive protocol \dref{cons2} does ensure the ultimate boundedness of both
the consensus error $\xi$ and the adaptive gains $d_i$. That is, the adaptive protocol \dref{cons2}
is robust in the presence of external disturbances.
The upper bound of the consensus error $\xi$ as given in \dref{upperbound}
depends on the dynamics of each agent, the communication graph, the upper bounds of the disturbances, and the parameters $\varphi_i$.
We should choose appropriately small $\varphi_i$ to get an acceptable upper bound of $\xi$.
\end{remark}

\begin{remark}
Compared to the robust adaptive protocol in \cite{19} which are only applicable to the case with matching disturbances,
the adaptive protocol \dref{cons2} works for general external disturbances.
This is a favorable consequence of introducing novel additive functions $\rho_i$, rather than multiplicative ones as in
\cite{19},  into \dref{cons2}. It is worth noting that the procedures in showing Theorem 2 is quite different
from those in \cite{19}.
\end{remark}
%\begin{remark}With the well-known $\sigma$ modification technique, the ultimate boundedness of the consensus error $\xi$ and the adaptive gains $c_i$ can be guaranteed. However, we make modifications on the adaptive laws into (13) to ensure $c_i\geq1$ so that we can explicitly get the upper bound of the consensus error $\xi$ which is shown in the proof.
%\end{remark}

\section{Simulation}
Consider a network of second-order integrators, described by (1), with
$$x_i=\left[\begin{array}{c}
x_{i1}\\
x_{i2}
\end{array}
\right], ~ A=\left[\begin{array}{c}
0\quad 1\\
0\quad 0
\end{array}
\right], ~B=\left[\begin{array}{c}
0\\
1
\end{array}
\right].$$
The communication graph is given as in Fig. 1, which clearly satisfies Assumption 1.
\begin{figure}[H]
\begin{center}
\includegraphics[width=2.2in]{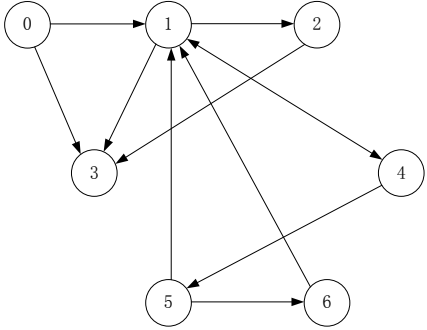}    % The printed column
\caption{The leader-follower directed communication graph.}  % width is 8.4 cm.
\label{fig1}                                 % Size the figures
\end{center}                                 % accordingly.
\end{figure}

Solving the LMI \dref{alg1} by using the LMI toolbox of Matlab gives a feasible solution matrix
$P=\begin{bmatrix}
1.7559 & -0.5853\\
-0.5853& 0.5853
\end{bmatrix}$. Then, the feedback gain matrices of \dref{cons1} are given by
$$K=\begin{bmatrix}-0.8543 &  -2.5628\end{bmatrix},\quad \Gamma=\begin{bmatrix}0.7298  &  2.1893\\
    2.1893  &  6.5678\end{bmatrix}.$$
Let $c_i(0)=1$, $i=1,\cdots,6$. The consensus errors $x_i-x_0$, $i=1,\cdots,6,$ of
the second-order integrators under the adaptive protocol \dref{cons1} are depicted in Fig. 2
and the adaptive coupling weights $c_i$ are shown in Fig. 3.
\begin{figure}[H]
  \centering
  \includegraphics[width=2.3in]{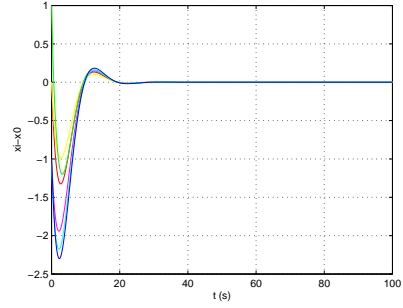}\\
  \caption{The consensus errors $x_i-x_0$, $i=1,\cdots,6$.}\label{Fig.2}
\end{figure}
\begin{figure}[H]
\centering
\includegraphics[width=2.3in]{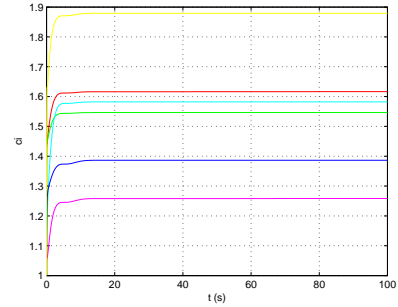}\\
\caption{The adaptive coupling weights $c_i$.}\label{Fig.3}
\end{figure}

Further, consider the case where the second-order integrators are perturbed by external disturbances.
For illustration, the disturbances associated with the agents are assumed to be
$\omega_0=\left[\begin{array}{c}
0.1\sin(2t)\\
0.3\sin(4t)
\end{array}
\right]$, $\omega_1=\left[\begin{array}{c}
0.2\sin(3.5t)\\
0.3\cos(2.5t)
\end{array}
\right]$, $\omega_2=\left[\begin{array}{c}
0.15\cos(4t)\\
0.2\sin(5t)
\end{array}
\right]$, $\omega_3=\left[\begin{array}{c}
0.3\sin(x_{32})\\
0.6\sin(3t)
\end{array}
\right]$, $\omega_4=\left[\begin{array}{c}
0.3e^{-2t}\\
0.15\cos(3t)
\end{array}
\right]$, $\omega_5=\left[\begin{array}{c}
0.2\sin(4t)\\
0.25\cos(3t)
\end{array}
\right]$, $\omega_6=\left[\begin{array}{c}
0.3\sin(5t)\\
\frac{0.4}{x_{61}^2+1}
\end{array}
\right]$, and the control input of the leader is assumed to be $u_0=e^{-0.1t}$.
Solving the LMI \dref{alg2} with $\varepsilon=2$ gives $Q=\begin{bmatrix}
0.2622  & -0.3517\\
   -0.3517  &  0.7395
\end{bmatrix}$ and then
$K=\begin{bmatrix}-5.0141  & -3.7372\end{bmatrix}$, $\Gamma=\begin{bmatrix}25.1412 &  18.7386\\
   18.7386  & 13.9665\end{bmatrix}$.
In \dref{cons2}, let $\varphi_i=0.1$ and $d_i(0)=1.5$, $i=1,\cdots,6$.
The consensus errors $x_i-x_0$, $i=1,\cdots,6,$ under the robust adaptive protocol \dref{cons2}
are depicted in Fig.4 and the coupling weights $d_i$ are shown in Fig. 5, both of which are obviously bounded.
\begin{figure}[H]
  \centering
  \includegraphics[width=2.3in]{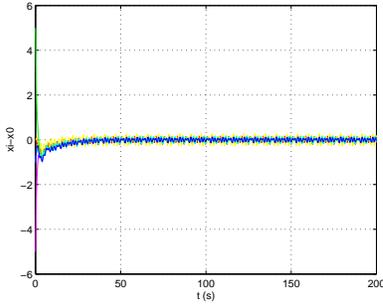}\\
  \caption{The consensus errors under the robust adaptive protocol \dref{cons2}.}\label{Fig.4}
\end{figure}
\begin{figure}[H]
\centering
\includegraphics[width=2.3in]{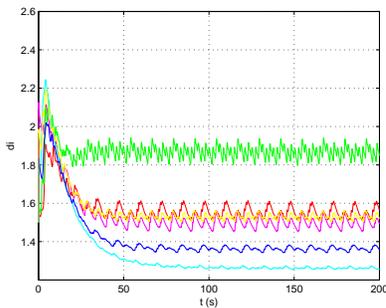}\\
\caption{The coupling weights $d_i$ in the presence of disturbances.}\label{Fig.5}
\end{figure}

\section{Conclusion}
In this paper, we have presented novel distributed adaptive consensus protocols for linear multi-agent systems with external disturbances
and directed graphs containing a directed spanning tree with the leader as the root. The adaptive consensus protocols, depending on only the agent dynamics an the relative state information of neighboring agents, can be designed and implemented in a fully distributed way.
One contribution of this paper is that the new distributed adaptive protocol is robust in the presence of general bounded external disturbances.
An interesting topic for future investigation is to design fully distributed adaptive protocols for nonlinear multi-agent systems
or the case with local output information of each agent and its neighbors.

\end{document}